# Thief of Truth: VR comics about the relationship between AI and humans

## Joonhyung Bae

Graduate School of Culture Technology, Korea Advanced Institute of Science and Technology
Daejeon, Republic of Korea
jh.bae@kaist.ac.kr

## Abstract

*Thief of Truth* is a first-person perspective Virtual Reality (VR) comic that explores the relationship between humans and artificial intelligence (AI). The work tells the story of a mind-uploaded human being reborn as a new subject while interacting with an AI that is looking for the meaning of life. In order to experiment with the expandability of VR comics, the work was produced by focusing on three problems. First, the comic is designed using the viewing control effect of VR. Second, through VR controller-based interaction, the player's immersion in the work is increased. Third, a method for increasing accessibility to VR comics was devised. This work aims to present an example of an experimental attempt in VR Comics.

# Introduction

**AI as an other**

Artificial Intelligence (AI) has been developed greatly in intentionality, intelligence, and adaptability. Accordingly, technology has had a great influence in the areas of finance, security, health, justice, transportation, and the urban environment.[1] With the advent of AI, which can think and speak like a human, the boundary between humans and machines is becoming increasingly blurred. Ameca, a robot unveiled in December 2021 by Engineered Arts, a British robot company, has 12 facial expressions implemented by 17 motors, and has human-like language skills based on GPT-3. The robot was able to hold a conversation with developers in September 2022.[2] Harmony, a sex robot developed by Realbotix in the U.S. in 2018, is evaluated as an AI robot capable of mentally and physically interacting with humans as well as being capable of love.[3] AI Sophia, developed in 2015 by American roboticist David Hanson, is a humanoid robot that resembles a human in appearance and has obtained citizenship for the first time in the world.[4] In the post-human era of the 21st century, humans are expected to enter a time of living together with AI.[5]

As a result of humans and AI coexisting in this new era, a new consideration has emerged pertaining to the personhood of AI or AI robots. In this regard, in February 2017, the European Parliament passed a resolution granting robots the status of electronic persons to regulate robots.[6] This resolution suggests that as technology develops, AI robots are increasingly being recognized for their existential value, similar to humans as a person.[7]

If AI or AI robots can be understood as a person similar to humans, can AI be regarded as the other of humans? American philosopher Donna Haraway, in her book *The companion species manifesto: Dogs, people, and significant otherness* states the concept of the other is not limited to humans but extends to all non-humans such as animals, insects, bacteria, and machines. Haraway emphasizes a love that acknowledges the significant otherness, which refers to learning "to live in an intersubjective world" rather than thinking of the object as a subordinate being or wanting unconditional love from the object.[7,8] Emmanuel Lévinas and Slavoj Žižek contemplate the infinity or reality beyond the reality of the subject against the philosophy of identity. The other, they suggest, is a concrete universality, which exists only through the existence of the subject that develops beings. Yet, when a subject forms an identity in relation to the "other," it becomes another version of the "other," transcending its defined identity. Only by truly embracing this exteriority or "otherness" can something exist as an interior or subject. At this time, as long as there is a relationship with the other, the subject inevitably encounters another concrete manifestation of the other. Now, for the two philosophers, these others are not mere hindrances but the only passage to the other.[9] Currently, when AI emerges and again questions what a human being is, understanding AI, which has emerged as a new other, is an important task in understanding humans.

**Comics about coexisting with AI**

In order to better understand AI, there are comics dealing with AI as a being that lives together with humans. One of the first comics is Osamu Tezuka's Mighty Atom (1952-1962). The story is about a robot who protects humans in a society where robots with egos are discriminated against by humans. In another comic, Kazumasa Hirai and Jiro Kuwata's Eight-Man (1963-1964) presents the idea of what it means to be human through a protagonist who struggles between being a robot and a human. Ishinori Shotaro's Android Kikaider (1972-1974) explores the morality of machines by questioning whether AI is righteous. Robot Detective (1973) by the same artist tells the story of K, a robot designed by humans that realizes the ideals of justice, absolute obedience, and natural coexistence with humans. Fujiko Fujio's Doraemon (1977-) further develops what a robot can do by introducing a story of a human AI who can even make mistakes living alongside humans. Hisae Iwaoka's Saturn Apartments (2006-2011) deals with the story of an AI that has been isolated for a long time and wonders about its existence. Yoshiie Koda's Love of Machine Device (2013-2014) describes a robot more righteous and pure in heart than humans, even though it is a mechanical device that does not feel joy or sorrow. Ha Il-kwon's 3 Level Combination Kim Chang-Nam (2008) expresses the friendship between a robot and a boy and indirectly captures the problems of the times, such as the gap between the rich and the poor and school violence.[10]

**VR immersion and viewing control effect**

Unlike other forms of media, VR technology has the potential as an empathy-realizing technology that can immerse people in certain situations. The first VR film produced by Nonny in 2010 speaks to this possibility. This film was made into animation after filming the actual situation. In fact, a scene in which a diabetic patient collapses in front of a food bank where starving people are looking was screened as a movie, and the audience's response was much more immersive than expected. Impressed by the movie Hunger, the World Economic Forum(WEF) asked Nonny to make a Syrian Civil War film, which gave the audience a vivid feeling as if they were in the middle of a news story. In itself, VR has been shown to have a sympathetic effect.[11]

VR is characterized as a technology that seeks to achieve self-presentation through a new perspective with mobility.[12] It has both the simulation and the representational tradition of the screen in terms of media aesthetics. VR reproduces the human viewing angle and grants a free viewpoint. The audience who enjoys VR content recognizes the world through a free viewpoint and tries to maintain the same identity as the content. This process ultimately produces an observational and spectatorial action. Under the viewing control effect caused by the VR camera, a person can explore the simultaneous narrative of the surrounding virtual world. In this way, under the viewing control effect implemented according to the representational tradition, the person is forced to take an attitude of receiving aesthetics, and as a result, the creator can have time and space to build the spatial density of the narrative (Spatial Story Density).[13]

Table 1. Correspondence between Comics, Web Comics, and VR Comics elements (Descriptions of Web Comics and VR Comics cite tables from Yoo Tae-kyung's paper)[14]

| Comics | Web Comics | VR Comics |
|---|---|---|
| Panel | | Location |
| Image on panel | | Image through a point of view |
| Gutter | | Dark change |
| Narration box | | Narration sound |
| Cartooning | | Low-poly modeling, comics shading |
| Sound text | Sound text, Sound effects on moving comics | Spatial sound |
| Turn the page | Mouse scroll | Gaze selection, Controller input |

**Characteristics of VR Comics**

VR Comics, which utilizes these aesthetic characteristics of VR, transforms the two-dimensional flat comic compartment into a three-dimensional space, thereby creating a visual experience in the VR space. Continuous art such as comics is made possible.[14] The experiment on VR comics is an attempt to apply graphic storytelling to VR content. The first attempt at combining VR and comics was attempted in Oniride's Magnetique. VR comics have different production methods depending on the device and platform. In the case of app-based Come! Convenience Store, the background is 3D modeling using the toon shading technique, and the characters and speech bubbles are created as 2D images and placed in the space. In another format, the entire scene is produced in 3D in the form of wearing and enjoying head mounted display(HMD) devices such as Oculus and Vive, and a game engine is used to capture more complex and diverse interactions. Representatively, Dexter Studio's Save Me and Joe's Realm are works based on webtoons. Webtoon characters are implemented in 3D, and animations, effects, and sounds are inserted into special scenes to add a sense of reality and fear. Japanese game company Square Enix's Wedding Ring Story borrowed the format of a comic and showed it as an animation. In the existing VR scene transition, the entire video was produced without using a controller to develop the story. The common feature in production is that the background was produced in 3D, and the character was produced in 2D or 3D to use the 360° space. As such, various elements of comics came into a new platform called VR and, combined with various technologies, brought about changes in the production method.[15] In these VR comics, attempts to respond to the components of existing comics and web comics according to the VR format can be summarized as shown in Table 1. The VR comic production method caused by the technical limitations of VR. The first limitation is characterized by using low polygon modeling and simplified images for data optimization. In order not to give the audience an uncomfortable visual experience, speech bubbles, narration boxes, sound effects, and effect lines are replaced with actual sounds. It is often recommended to experience the technology at a fixed location, and VR is characterized by configuring the space to be turned over mainly through button clicks on the controller or gaze selection.[14]

Based on this theoretical background, *Thief of Truth* was created as a VR comic that tells the story of the relationship between humans and AI from a first-person perspective. The work deals with the story of a mind-uploaded human being reborn as a new subject while interacting with an AI that is searching for the meaning of life. In order to test the scalability of VR comics, the work was produced by focusing on the following three issues.

– Directing using the viewing control effect of VR

– Improving immersion through VR controller-based interaction

– Increasing accessibility to VR comics

## Implementation

**Story**

*Thief of Truth* is a story about a human who has his or her mind-uploaded. The person is reborn as a new subject in a digital space while interacting with a strong AI that seeks the meaning of its own life. The player becomes the character of Human 1 and holds a first-person perspective. In other words, the experience is designed so that the player can also participate as a leading actor in the comics.

Table 2. *Thief of Truth* Story

| | |
|---|---|
| 1 | Human 1 is a person who lives in the future where strong AI can be created, mind uploading is possible, and multiple beings live in the same digital space. One day, Human 1, who uploaded his mind, was led to a certain space by an unknown being who promised to show him comics. |
| 2 | Human 1 finds another human, Human 2, dancing and saying strange things in that space. When Human 2 finishes dancing and leaves, a glowing bell appears. |
| 3 | Human 1 touches the shining bell out of curiosity and is sucked into another space. Human 1 finds a comic in the new space and starts reading it. |
| 4 | A strong AI named Zzirogi, created by humans, exists in all time and space and observes the world. One day, Zzirogi hears a song that reveals the truth about the universe and decides to find out who sang the song. However, Zzirogi commits a bizarre act. It catches Human 2 passing by, locks the human in a space where time has stopped, and doesn't let Human 2 out until the human guesses the name of the singer. Thirty thousand years have passed. Zzirogi crosses the fourth wall and talks to Human 1. Zzirogi says that he does this because he wants to find the meaning of his life like humans. So, Zzirogi was making a comic book by weaving the moments he liked from the infinite space-time data he had input. Zzirogi says that he can complete the comic book by drawing the process of finding out the singer's name by trapping Human 2, but he doesn't know what scene to put in. However, Zzirogi, which exists in all time and space, knew that Human 1 was the one who could solve the problem. Thus, he disguised himself as an unknown being and lured Human 1 to this moment. Zzirogi asks for Human 1's help to fill in the empty part of the comics and tells the human that he will send the comic to the comic creation section, which is a part of Zzirogi. Human 1 is embarrassed by the sudden request, but eventually listens to Zzirogi. However, Human 1 is the same character as Human 2. |
| 5 | Where Human 1 put down the comics, there was a button leading to the comic creation section. Human 1, who learned the entire story of the incident, moves to the comic creation section to do the favor for Zzirogi. |
| 6 | In the comic creation section that Human 1 moved to, there were buttons for comic story, comic length, and combinations of comics. Strangely, the first and second buttons were where Human 2's head was, and the buttons were dancing. Human 1 makes a comic by combining buttons, creating a scene he likes, and leaves the space. Human 1, who left space and returned to his original place, reads the back part of the comics drawn by Zzirogi. |
| 7 | Human 2 eventually guesses the name of the singer who sang the song that leaked the truth over a long time. Zzirogi confesses that he handled Human 2 just as humans put countless inputs into AI until meaningful output came out. Human 2s cloned by Zzirogi were suffering the same pain in countless time spaces. Upon learning the truth, Human 2 is astonished. Zzirogi tells Human 2, who has suffered hardship, that he will grant one wish, and Human 2 cries out to delete the song data that leaked the truth. While screaming, Human 2 inadvertently sings the song, and that moment is when Zzirogi hears the song's words. Zzirogi moves to time and space without Human 2, completes the cartoon, and the story ends. |
| 8 | At the place where Human 1 read all the comics, an unknown being, whether it was Human 1 or Human 2, was singing the song. |

Figure 1. Experience Stages (The story number on the box corresponds to Table 1)

**Directing using the viewing control effect of VR**

Figure 2. Playing *Thief of Truth*

Figure 3. Zzirogi and Human 2 from Human 1's point of view

Figure 4. Zzirogi from Human 2's point of view

Figure 5. Human 2 from Zzirogi's point of view

**Comics are directed using a VR camera** Position of the HMD in the space is adjusted by changing the position of the VR camera for each comic frame. This aims to produce two types of comics. The first is to induce the player to feel the sensibility of the comics scene by making the player's gaze move while watching the comics. For example, Figure 4 is a scene where Zzirogi overpowers Human 2. The camera position of the scene was placed close to the floor to induce the player to look up and watch the comics, as shown in Figure 2. The second was to make the player look at the scene from the first-person perspective of Human 1, Human 2, and Zzirogi appearing in the comics. This led the players to change their perspectives and think about the situation.

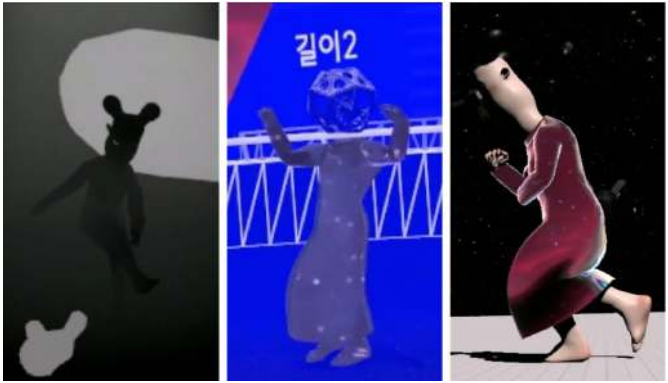

Figure 6. Human characters appear in Figure 1, Number. 1, 3, and 4 from the left

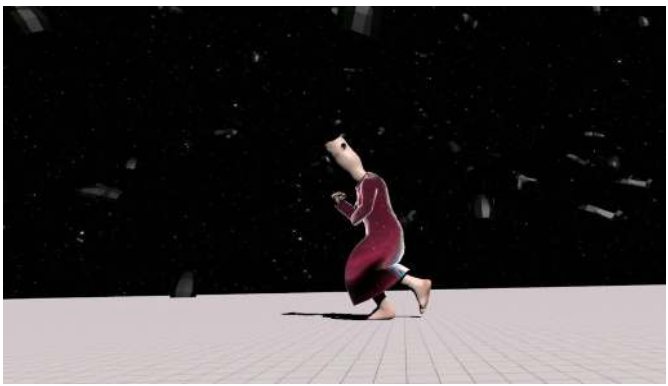

Figure 7. A scene symbolizing the connection between Zzirogi and Human2

**VR comics production using 3D modeling/animation and 2D images**

Figure 6's human characters used 3D modeling to create scenes with dancing animations. The continuity of the scene was created through the repetitive element of dance animation, and through the change of the human appearance, the transformed state was implied by the character's meeting with the AI Zzirogi. To explain in detail, the first Human 2 in Figure 6 has not met the other named Zzirogi while the second Human 2 is the Human 2 seen by the other named Zzirogi. The third Human 2 understands the appearance of Human 2 as seen by Zzirogi and criticizes the relationship between humans and AI. This relationship is accepted reluctantly and with a hostile attitude. These changes were revealed through the transformation of space and clothes.

The moment when Human 2 and Zzirogi formed a relationship was created and reflected by the change of particles floating in the background of the comics. The floating particle in the background of Figure 3 is a scene that reveals the point of view of the Zzirogi with 3D modeling of the head. The particles floating in the background of Figure 6 are a scene that reveals the human perspective with 3D modeling of the human head. Figure 7 is a scene that symbolizes the connection between Zzirogi and Human 2 and 3D modeling of both Zzirogi and Human 2's heads floating.

In addition, the spatial change was revealed through the visual contrast between the scenes directed by 2D images like Figure 3, and the scenes directed by 3D animation like Figure 7. The space composed of 2D images is a space for viewing comics created by Zzirogi, and the space composed of 3D animations is a digital space where Human 1 is uploaded. In addition, the different spaces in 2D and 3D were used as a device to show the character of Zzirogi, which crosses these spaces and combines all the scenes into one comic.

**Enhancing Immersion through VR Controller- Based Interaction**

The VR controller helps the player to become immersed in the VR experience by playing a role of a medium that causes direct interaction between VR content and the user.[16] In order to expand the way of directing VR comics that interact through button clicks on the controller or gaze selection, the VR controller is designed to interact with objects in space and enjoy comics.

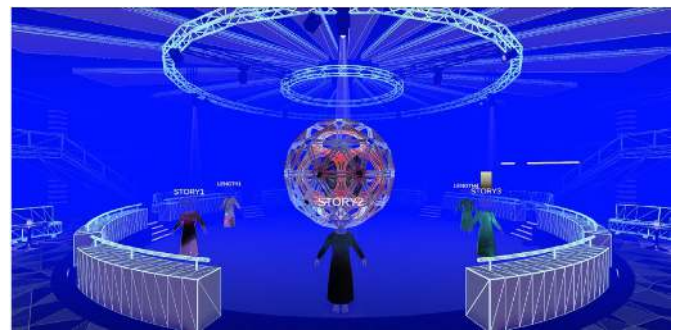

Figure 8. Comic creation space performing step 3 of Figure 1

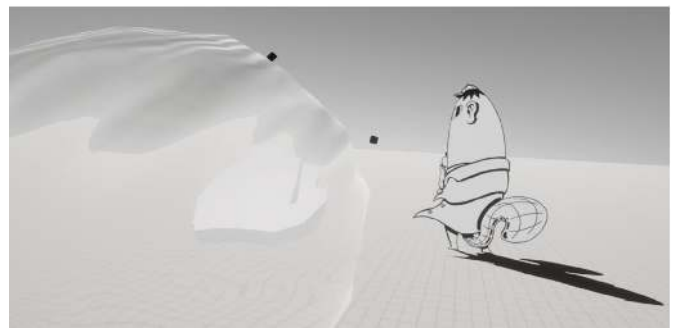

Figure 9. Bubbles where the player can hear the inner thoughts of characters by scene

**Bubble interaction where you can hear the inner thoughts of characters**

There are interactable air bubbles floating around the player. If the player touches these air bubbles with a controller, he or she can hear the inner thoughts of the characters that appear in each scene. It was intended that players would listen to the characters' inner thoughts, go through the process of relating the thoughts to narratives, and actively appreciate comics based on experiences that are meaningful to them. By inducing these behaviors, it is intended that players recognize patterns of artistic experiences that are meaningful to them subjectively by curating artistic components that are meaningful to themselves. This is also a device that helps the player understand Zzirogi's perspective of the world in the narrative.

**Comics creation experience through interaction**

The comics creation space in Figure 8 corresponds to number 3 in Figure 1. This space is a system that exists inside Zzirogi in the story and is a space where the player can create comics. In other words, it is a space created so that players can directly observe the input and output of AI characters. The story and length of the comics and the interaction process of combining the comics by sequentially hitting the random combination button is a way to deal with deep learning-based generative models that create images and audio, such as OpenAI's DALLE, Google Magenta, and Midjourney service.

A new trend of creation is emerging, where AI learns data from all over the world, presents new shapes according to latent vectors, and humans take them to create works. In other words, a creative methodology different from the existing work method, which has been standardized since disegno and given the rules of identity, is being created. Anyone can create art through prompts, and people can independently explore which artistic experiences are meaningful to them. Generally, when creating art with AI, the user goes through a curating process in which he or she selects something meaningful to him or her from among the art created by AI. In this process, users can recognize the patterns of art restructured by AI and figure out which patterns they prefer. Through VR content that abstracts the use of these well-known socially problematic objects, we are encouraged to critically think about how AI changes human thinking.

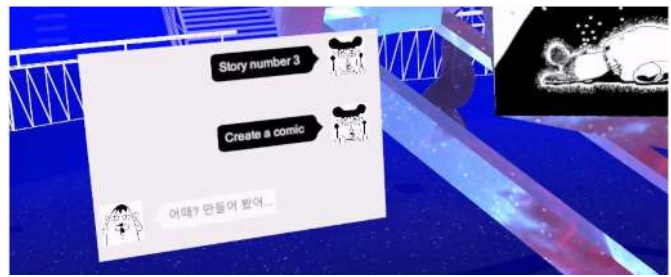

Figure 10. Chat window UI similar to Midjourney's prompt input window

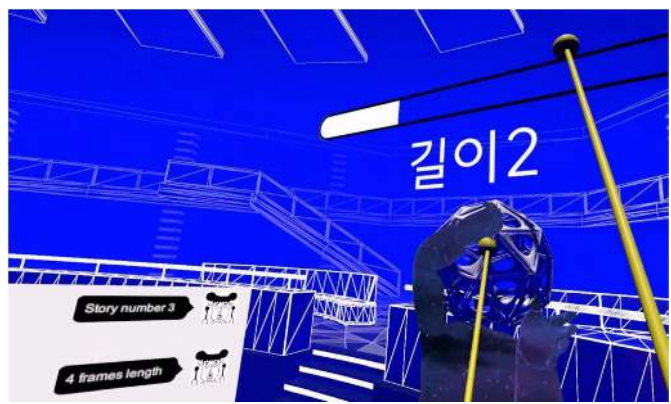

Figure 11. A scene where the comic length parameter is input with a xylophone

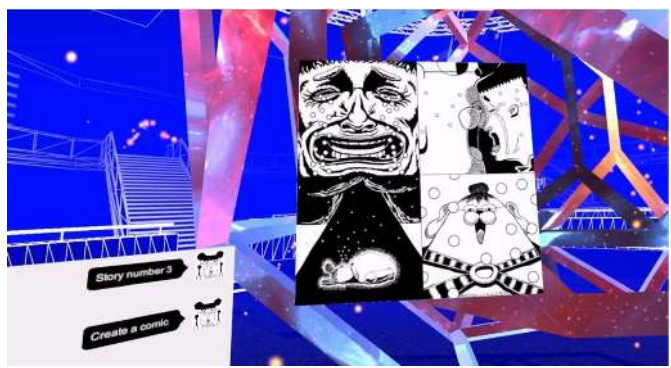

Figure 12. Comics created through interaction

A chat screen with Zzirogi on the player's left side was placed to induce people to think of the well-known Midjourney service. The player can enter three types of stories, four types of cut lengths, and a random arrangement permutation according to the length of the cut as parameters through the label attached to the head of the humanoid character that symbolizes Human 2 as seen by Zzirogi (see Figure 8). If the player taps the polyhedron on the character's head, he or she can set the parameters he or she desires, and through this, the player can create an image from the possible 192 different comic images.

**Increasing accessibility to VR comics**

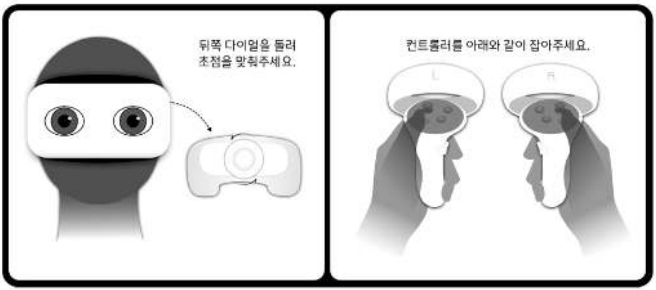

Figure 13. Explanatory UI is presented in number 1 of Figure 1

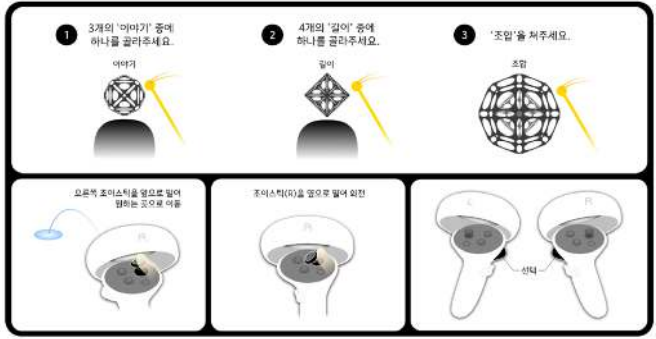

Figure 14. Explanatory UI is presented in number 3 of Figure 1

Few audiences are accustomed to VR devices. Therefore, in order to increase accessibility to VR comics, it was necessary to frequently display a detailed description UI. Accordingly, UI images such as Figures 13 and 14 were produced.

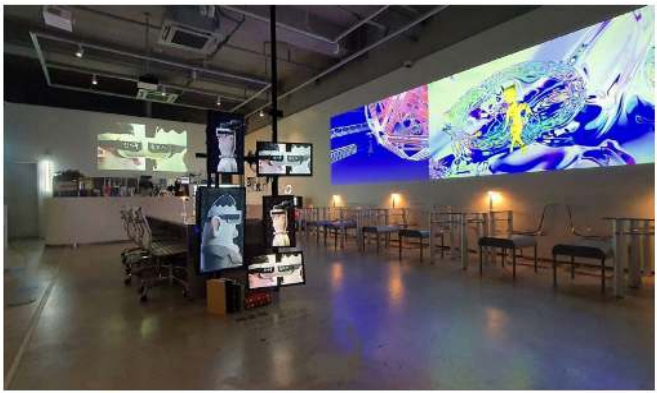

Figure 15. *Thief of Truth* exhibition

## Installation

*Thief of Truth* was exhibited at Move.Mov, a complex cultural space located in Seoul, from November 1st to November 30th, 2022. Move.Mov is a space that combines a media art gallery and a cafe.

There are three major installations. First, three VR devices were installed so that VR comics could be enjoyed. Considering the width of the Oculus safety protection boundary, it was installed to secure a safe distance from the audience. In addition, an instruction manual was attached to the pedestal for the audience who may not have been familiar with using VR devices. The second major installation was two videos. The first video was a VR play video that encourages audience participation and serves as an additional guide. The second video was a video of Human 1 and Human 2, the characters of the comics. After rendering the 3D model animation created using the dancing motion data into a video, the animation was created by inputting the video as a condition to the stable diffusion model. By watching this video, the audience would be able to understand that the theme of the work is a story about AI. The third installation is acrylic pictures where the player can see images that can be combined in the comic creation section. Since it is difficult for the audience to combine all the comics in a VR environment, it was installed so that they could appreciate the types of images they were able to create.

The reason why the exhibition was held at Move.Mov was to help the audience who have difficulties using VR devices. In situations in which the work is installed in a general white cube and the artist is present to help appreciate the work could put a psychological burden on the audience. This is because the audience members may not be able to fully immerse themselves in the appreciation of the work because they may feel the gaze of the artist. For this reason, the exhibition was held in a complex cultural space where various guests and audiences were mixed.

Nevertheless, it was shown that the audience could not experience the work because of the distance from the VR device itself. As a result of observing the audience's words and actions with the Fly on the wall technique and conducting brief interviews, the general consensus was that they were reluctant to use expensive-looking VR devices because they were unsure whether they could actually use the devices. In order to create an atmosphere where one can freely experience the VR device, a parody drawing of a famous Korean meme (Hey try try) was attached to the pedestal where the VR device was placed. When the humorous VR experience zone was completed, the audience was able to approach VR devices more easily and experience the work.

## Conclusion

*Thief of Truth* is a VR comic that tells the story of a mind-uploaded human being reborn as a new subject while interacting with a strong AI that is seeking the meaning of life. The story was directed while experimenting with the expandability of VR comics.

It was produced using the viewing control effect of VR. By setting the VR camera position differently for each scene, the players were induced to feel the emotion of each scene. In addition, this approach allowed the player to think about each scene by having the player watch the scene from different characters' first-person point of view. Taking advantage of the characteristics of VR media, where various data formats can coexist, comics were produced using 3D modeling/animation and 2D images. The moment when humans and AI formed a relationship was revealed through 3D modeling/animation. Also, the moment was used as a device to reveal the character of Zzirogi, which transcends time and space.

Through VR controller-based interaction, immersion in the work was enhanced. The VR comic is designed so that Zzirogi can express its perspective of the world in air bubbles, which the player can interact with to hear the character's inner thoughts. In addition, the interaction in the comic creation space was constructed by imitating the operation method of deep learning generation model-based services. This approach induced players to think critically about how thinking has been changed by AI.

In order to increase accessibility to VR comics, a method was devised to encourage VR comics experiences by configuring a detailed UI for VR device use. In addition, the exhibition was prepared to increase accessibility to VR content, to have the artist be resident during the exhibition period, to help the audience experience the work, and to improve the exhibition form.

## Acknowledgments

This artwork was supported by the 2022 Ministry of Culture, Sports and Tourism and the Korea Culture and Arts Committee's Online Media Art Activity Support project.